\documentclass[pre,twocolumn]{revtex4}

\usepackage{graphicx}

\begin{document}

\title{Analytic continuation of QMC data with a sign problem}
\author{A.\ Macridin} \affiliation{University of Cincinnati,
  Cincinnati, Ohio, 45221, USA} 
\author{S.  P. \ Doluweera}
\affiliation{University of Cincinnati, Cincinnati, Ohio, 45221, USA}
\author{M. Jarrell} \affiliation{University of Cincinnati, Cincinnati,
  Ohio, 45221, USA} 
\author{Th.\ Maier} \affiliation{Computer Science
  and Mathematics Division, Oak Ridge National Laboratory,\ Oak Ridge,
  Tennessee 37831, USA }

\date{\today}

\begin{abstract}
  We present a Maximum Entropy method (MEM) for obtaining dynamical spectra
  from Quantum Monte Carlo data which have a sign problem. By relating
  the sign fluctuations to the norm of the spectra, our method 
  properly treats the correlations between the measured quantities and the
  sign.  The method greatly improves the quality and the resolution of
  the spectra, enabling it to produce good spectra even for poorly
  conditioned data where standard MEM fails.
\end{abstract}

\maketitle

\section{Introduction}

One of the greatest advantages of Quantum Monte Carlo (QMC) simulations is the
possibility to deal with complex and large size systems.  The tremendous
increase in computing capabilities and the development of new QMC based
algorithms in recent years gives rise to new opportunities for QMC simulations.
However, the possibility of producing new data implies a series of new
problems for processing and analyzing the data.

Despite the QMC successes, these simulations have some general limitations.
One such difficulty is the sign problem, which affects a large class of 
quantum models and appears when the sampling weight of some configurations 
is not positive definite.  Another drawback of QMC simulations is that, while 
static measurements are easily obtained, calculating dynamical quantities 
is extremely difficult.  When the sign problem is present, this difficulty is 
much more serious. In the past, the limited computing capabilities available 
didn't allow for simulations with a small average sign. With the advent of 
new parallel vector machines such as the CRAY X1 at ORNL however, the speed 
of these calculations is significantly improved, making simulations with a 
small average sign feasible. This necessitates major improvements in the 
methods used to analyze the new data. 

The standard technique of extracting dynamical spectra from QMC
simulations based on the Matsubara-time path integral formalism
is the Maximum Entropy Method (MEM)~\cite{mem,mem1,mem2,mem3}.  The  
dynamical properties contain important information about excited 
states and describe the system's response to different external 
perturbations, making the direct connection between model and  
experiment.  Therefore an  algorithm able to produce dynamical 
quantities is of crucial importance. The goal of this paper is to 
describe an improved MEM technique of calculating dynamical 
properties of a system from QMC data with a sign problem.

MEM recasts the problem of spectra calculation from a deterministic
problem to one of probability optimization. In principle, by knowing 
the imaginary time response functions, the dynamical spectra can be  
obtained by solving an integral equation.  However, in practice, the
calculation of spectra is an ill-defined problem. Due to the fact 
that QMC provides information on a finite number of time points 
with a certain error bar, an infinite number of solutions consistent 
with the data exists.  MEM is an algorithm which, based on  Bayesian 
inference~\cite{bayesian}, provides the most probable spectrum 
compatible with the available data~\cite{membay}.

The spectrum probability is calculated assuming {\em Gaussianly 
distributed} and {\em uncorrelated} data.  As the central limit 
theorem requires, the statistics of any average is Gaussian as long 
as the average is taken over a large number of uncorrelated points. 
Methods have been developed to reduce the correlations in the data, 
both those between adjacent measurements and those at different 
Matsubara time points in the same measurement~\cite{mem}.

However, when the sign problem is  present, the QMC data becomes 
very poorly conditioned, which greatly complicates the MEM problem. 
Non-Gaussian distributions  and  strong correlations of the data turn out 
to be very severe problems. They cannot be removed by the standard
techniques, mainly due to the strong correlation between the data 
and the averaged sign of the configurations which produce these data. 
This makes it essentially impossible to calculate spectra long before 
the minus sign problem makes the calculation of static properties
impractical.  In this paper we address this problem and describe a
solution which greatly increases the resolution of MEM
when calculating spectra from such poorly conditioned data.

This paper is organized as follows. In Sec.~\ref{sec:mem} we introduce the
general MEM formalism. In Sec.~\ref{sec:datas} we discuss and exemplify the
problems which appear when the QMC simulations suffer by the sign problem. A
solution to the problem is given in  Sec.~\ref{sec:newmem}. A comparison of
spectra obtained with the standard and the improved method is presented in
Sec.~\ref{sec:newdata}.  The conclusions are given in Sec.~\ref{sec:conc}.

\section{MEM Formalism}
\label{sec:mem}
We start with a brief introduction of MEM\cite{mem}.  MEM is an
algorithm which aims to determine the spectral decomposition of
one- or two-particle Green's functions.  Most QMC methods only 
produce estimates of the imaginary time Green's functions.  The 
relation between the spectral density, $A(\omega)$, and the imaginary 
time Green's function, $G(\tau)$, is given by an integral
equation~\cite{textbooks}
\begin{equation}
\label{eq:inteq}
G(\tau)=\int K(\tau, \omega)A(\omega) d\omega 
\end{equation}
\noindent where the kernel,  $K(\tau, \omega)$, is given by
\begin{equation}
\label{eq:kernel1}
K(\tau,\omega)=\frac{e^{-\tau \omega}}{1+e^{-\beta \omega}}
\end{equation}
\noindent for the one-particle Green's function, and respectively
\begin{equation}
\label{eq:kernel2}
K(\tau,\omega)=\frac{1}{\pi} \frac{\omega e^{-\tau \omega}}{1-e^{-\beta \omega}}
\end{equation}
\noindent for the two-particle susceptibility~\footnote{The standard notation
for the two-particle susceptibility (spectrum) is  $\chi(\tau)$ (
$\chi''(\omega)/\omega$). In this paper we use  $G(\tau)$ ($A(\omega)$).}.

The determination of the spectrum is an ill-posed problem, since an
infinite number of solutions exists which are consistent with the QMC
data and associated error bars.  MEM selects from these solutions the 
most probable one. According to Bayesian logic~\cite{bayesian}, given 
the data $G$, the conditional probability of the spectrum $A$, $P[A|G]$, is 
given by
\begin{equation}
\label{eq:bayesian}
P[A|G]=P[G|A]~P[A]/P[G]\,.
\end{equation}
\noindent Here $P[G|A]$ is the {\em likelihood function} which represents 
the conditional probability of the data $G$ given $A$, $P[A]$ is the {\em
prior probability} which contains prior information about $A$ and
$P[G]$ is called the {\em evidence} and can be considered a
normalization constant.

The {\em  prior probability} is given by
\begin{equation}
\label{eq:prior}
P[A]=e^{\alpha S},
\end{equation}
\noindent with a real positive constant $\alpha$ and the entropy function $S$
defined by
\begin{equation}
\label{eq:entropy}
S=\int d\omega (A(\omega) -m(\omega)-A(\omega)\ln[A(\omega)/m(\omega)]).
\end{equation}
\noindent $m(\omega)$ is a function called ``default model''. The specific form
of the entropy function is a result of some general and reasonable assumption
imposed on the spectrum, like subset independence, coordinate invariance,
system independence and scaling. By defining the entropy relative to a default
model, the prior probability  is also used to incorporate  prior knowledge about
the spectrum, such as the high-frequency behavior and certain sum-rules. In the
absence of data the resultant spectrum will be identical to the model. The
entropic probability and its consequences are discussed at large in a series of
papers~\cite{entropy1, entropy2,entropy3,kangaroo,mem}, and does not constitute
the subject of this study.

The  main focus of this investigation is the calculation of the {\em likelihood 
function},  $P[G|A]$. The central limit theorem  shows that the distribution
of the data obtained in a QMC process is always Gaussian if every data point is 
taken as an average of a large enough number of measurements so that different 
data are independent. This implies
\begin{equation}
\label{eq:likely}
P[G|A]=e^{-\chi^2/2}
\end{equation}
\noindent where 
\begin{equation}
\label{eq:chisq}
\chi^2=\sum^L_{i,j=1}(\bar{G}_i-G_i(A))[C^{-1}]_{ij}(\bar{G}_j-G_j(A)),
\end{equation}

\begin{equation}
\label{eq:Gave}
\bar{G}_i=\frac{1}{M}\sum^M_{m=1} G^{(m)}_i
\end{equation}
\noindent with the covariance

\begin{equation}
\label{eq:cov}
C_{ij}=\frac{1}{M-1}\sum^M_{m=1} (\bar{G}_i-G^{(m)}_i)(\bar{G}_j-G^{(m)}_j).
\end{equation}

\noindent Here we considered that QMC provides $G(\tau)$ on $L$ time points
$\tau_i$, and denote $G(\tau_i) \equiv G_i$.  For every time point, $\tau_i$,
we have $M$ measured $G^{(m)}_i$, centered at $\bar{G}_i$ (Eq.~\ref{eq:Gave}).
$C$ in Eq.~(\ref{eq:cov}) is the covariance matrix which characterizes the
second moment of the data. $G_i(A)$ in Eq.~(\ref{eq:chisq}) is the value of
$G(\tau_i)$ which corresponds to the spectrum $A(\omega)$ according to
Eq.~\ref{eq:inteq} or its discretized form.

MEM requires Gaussianly distributed data. Otherwise, the likelihood
probability defined in Eq.~(\ref{eq:likely}) has no meaning. In theory, 
the requirement for a Gaussian distribution is achieved by averaging many
measurements to obtain one data point. However, in practice when
computational resources are limited, this condition is often 
difficult to satisfy.  In MEM literature the data points obtained 
by averaging  many measurements are called {\em bins}.
The usual way to remove the correlation
between bins is to re-average  (coarse-grain) more successive
bins which results in increasing the number of measurements per bin.
However, for a fixed amount of data, this process of increasing the
bin size will reduce the number of data points (bins). If the number
of bins is too small, the data cannot properly describe a statistic
process, and the covariance matrix becomes pathological. Therefore
a successful MEM for correlated data requires a large number of 
measurements, implying both large bins and a large number of bins.

The correlation of data between different time points, is the other
relevant problem which causes MEM to fail. These correlations can be
removed by a rotation $U$ which diagonalizes the covariance matrix.
\begin{equation}
\label{eq:covr}
C'_{ij}=(U^{-1}CU)_{ij}=\sigma^2_i \delta_{ij}.
\end{equation}

\noindent The data and the kernel should also be rotated accordingly
\begin{equation}
\label{eq:kerG}
K'=U^{-1}K,~~~~G'=U^{-1}G
\end{equation}
\noindent The rotated $G'_i$ are the statistically independent directions,
and in this basis, $\chi^2$ reduces to

\begin{equation}
\label{eq:chisqr}
\chi^2=\sum^L_{i=1}\frac{(\bar{G'}_i-\sum^{N_f}_{\nu=1}K'_{i\nu}A_\nu)^2}{\sigma^2_i}.
\end{equation}
\noindent In Eq.~\ref{eq:chisqr} the discretized form of Eq.~\ref{eq:inteq} was
used, the summation over $\nu$ meaning summation over frequency.

\section{Data produced by QMC simulations with sign problem}
\label{sec:datas}

\begin{figure}[t]
\centerline{ \includegraphics*[width=3.2in]{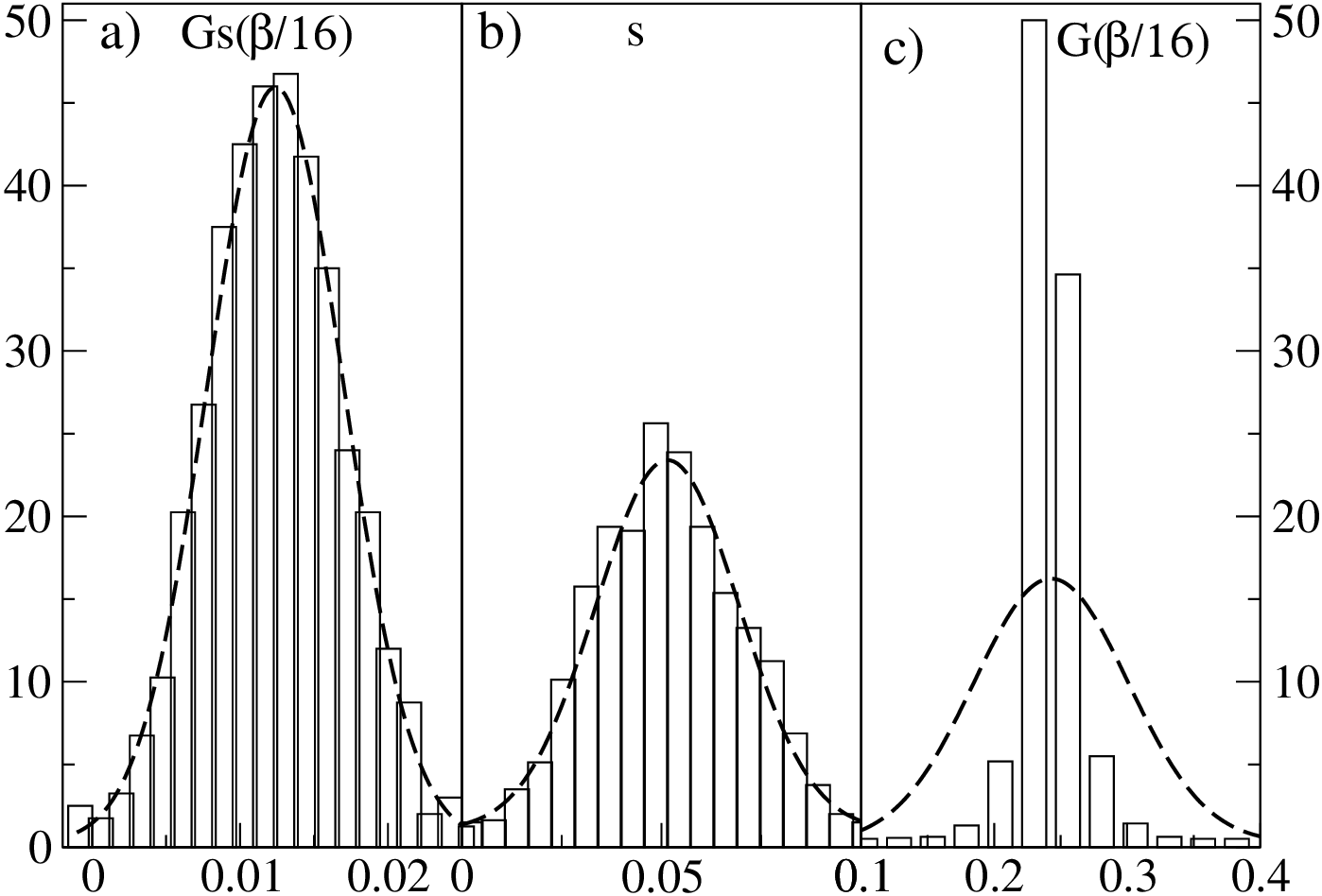}}
\caption{Histogram of the distribution of 
a) $Gs(\beta/16)$, b) sign, $s$ and c) $G(\beta/16)$ when the bin size is increased 
five times which corresponds to $3000$ measurements per bin.
The dashed lines represent the best  Gaussian fit to the data.}
\label{fig:hist}
\end{figure}

When the sign problem is present in QMC simulations, the condition of
Gaussianly distributed $G_i$ becomes more difficult to satisfy. Very often, a
huge number of measurements, beyond the available computing possibilities,
would be required to accomplish this task.

The difficulty in obtaining good data points for $G_i$ can be easily understood
from the measurement process.  In a QMC process where the sign of the sampling
weight is negative, it can no longer define a probability.  Therefore, the
sign of the sampling weight must be associated with the measurement. For the
Green's function, we can no longer measure $G_i$ but rather the product
of it and the sign $s$ of the configuration, $Gs_i \equiv
G_i*s$, and the sign  $s$. At the end of the simulation,
i.e.\ after a large number  of measurements, we then obtain
$\bar{G}_i=\overline{Gs_i}/\bar{s}$, where the overbar denotes averaging over
the  number of measurements.  Two problems related with the sign affect the quality of
the $G_i$ data. First, in order to obtain good data points $G_i^{(m)}$, for every 
data point we need to
average a very large number of $Gs_i$ and $s$ and afterwards
calculate $G_i^{(m)}=Gs_i^{(m)}/s^{(m)}$. 
Here $Gs_i^{(m)}$ and $s^{(m)}$ both denote averages over the
measurements that form the bin $m$.
Smaller average signs $\bar{s}$ worsen the problem, since any small variation of $s$
has a large effect on $G_i$ ($\propto 1/\bar{s}^2$). Second, as within the same 
bin $m$ there is a strong correlation between different data points $Gs_i^{(m)}$,
there is also a strong correlation between data points $Gs_i^{(m)}$ and $s^{(m)}$. The
points $G_i^{(m)}=Gs_i^{(m)}/s^{(m)}$
are obtained by a nonlinear operation of these correlated
quantities, and there is no reason to expect them to be normal distributed.

In order to exemplify the problems discussed above, we employed a QMC based
algorithm~\cite{dca} to produce a very large amount of data for the
single-particle Green's function and  the two-particle spin susceptibility in
the two-dimensional Hubbard model on a square lattice. The Hubbard model is
characterized by the single-particle hopping $t$ between nearest neighbors and
the on-site Coulomb repulsion $U$. We choose $t=0.25$ so that the bandwidth
$W=2$ and set $U=W$. To make the sign-problem worse, we add a next-nearest
neighbor hopping $t'=-0.3t$ to frustrate the lattice. We perform calculations
on a 16-site $4\times 4$ cluster at $15\%$ doping, down to temperatures
$T=0.125t$ where we experience a severe sign-problem, $\bar{s}=0.051$.  We
simulate the model using the dynamical cluster approximation (DCA) with the
Hirsch-Fye algorithm as a cluster solver.\cite{dca}  The DCA is a
coarse-graining approximation, in which the one particle Green's function is
coarse-grained in the first Brillouin zone of the reciprocal space of the
lattice.  It is defined over cluster points $K=(K_x,K_y)$ and imaginary time
$\tau$, and accurately describes short-ranged correlations.  We performed the
simulations on the Cray $X_1$ supercomputer at Oak Ridge National Laboratory to
cope with the large amount of data needed in simulations with small average
signs. We calculated $8000$ data points (bins)
 $(Gs_i^{(m)},s^{(m)})$, and for every data point we
averaged $600$ QMC measurements.

In Fig.~\ref{fig:hist} we show histograms of the data distribution 
when the bin size is increased five times, which corresponds to an average of 
$3000$ measurements per bin.
Both $Gs(\beta/16)$ (Fig.~\ref{fig:hist} (a)) and $s$
(Fig.~\ref{fig:hist} (b)) are normally distributed to a good
approximation, unlike $G(\beta/16)$ data points (Fig.~\ref{fig:hist}
(c)) which are strongly peaked, being characterized by a large positive
kurtosis~\cite{kurt}.  Similar distributions of data are observed (not
shown) for the other values of the imaginary time. In order to become
Gaussianly distributed the $G$  data require averaging over a much
larger   number  of measurements  than $Gs$ and $s$ data.  In our case
this number is about five times larger but this value is  dependent on
the specificity of the problem considered, being determined by both the
magnitude of the correlations and the value and the distribution of the
sign~\footnote{By rebinning $G$ we mean rebinning $Gs$ and $s$ and
afterwards obtaining $G$ as the ratio of these two quantities. Much
worse results are obtained if successive $G$ data points are rebinned.}.

The way to achieve good data for MEM is {\em i)} 
rebinning $(Gs_i,s)$ until they
become normal distributed, and {\em ii)} remove the correlations between data
points $Gs_i$ and $s$ by a rotations in the space $(Gs_i,s)$.  However, the
problem that arises is the calculation of $\chi^2$
(Eq.~\ref{eq:chisq},~\ref{eq:chisqr}) in this basis which now includes the extra sign
dimension. This issue will be discussed in the next section
(Sec.~\ref{sec:newmem}).

\section{Likelihood function}
\label{sec:newmem}

\subsection{Formalism}
\label{ssec:form}

Denoting $h \equiv (Gs,s)$, the likelihood function is defined as
$P[h|A]$, since the measured quantities in the QMC process are the $h$
points (and not $G$).  As we showed in the previous section, for
acceptable values of the bin size, the data $h$ are to a good
approximation Gaussianly distributed.  Therefore, the likelihood
function will have the same form as Eq.~\ref{eq:likely}, with $\chi^2$
\begin{equation}
\label{eq:chisqn}
\chi^2=\sum^{L+1}_{i,j=1}(\bar{h}_i-h_i(A))[C_{h}^{-1}]_{ij}(\bar{h}_j-h_j(A))\,.
\end{equation}

\noindent The covariance matrix has now the dimension $(L+1) \times (L+1)$,
\begin{equation}
\label{eq:covn}
{C_h}_{ij}=\frac{1}{M-1}\sum^M_{m=1} (\bar{h}_i-h^{(m)}_i)(\bar{h}_j-h^{(m)}_j).
\end{equation}

\noindent The only problem which remains to be solved is finding an equation
for $h(A)$, since Eq.~\ref{eq:inteq} only provides a relation for $G(A)$.
In order to achieve this we do the following: First we absorb the
sign into the spectrum, i.e. we define ${\cal{A}}$ as
\begin{equation}
\label{eq:an}
{\cal{A}}(\omega)=sA(\omega)~.
\end{equation}
\noindent Instead of searching for a spectrum $A$
which satisfies Eq.~\ref{eq:inteq} we search for ${\cal{A}}$
which satisfies
\begin{equation}
\label{eq:ks}
Gs(\tau)=\int K(\tau, \omega){\cal{A}}(\omega) d\omega~.
\end{equation}
\noindent Second, we consider the spectrum normalization sum-rule
\begin{equation}
\label{eq:sumruleg}
B=\int A(\omega) d\omega,
\end{equation}
\noindent which  implies
\begin{equation}
\label{eq:sumrules}
s =\int \frac{1}{B}{\cal{A}}(\omega) d\omega~.
\end{equation}
\noindent Here $B$ is a constant, equal to one
for the the one-particle spectra and
equal to $\chi(T)$ for the two-particle case~\footnote{Do not confuse
the static susceptibility, $\chi(T) \equiv\chi(\omega=0)$, with
$\chi$ defined in Eq.~\ref{eq:chisq}.}.
Because the sign $s$ was absorbed into the definition of ${\cal{A}}$
we relate the sign fluctuations  to the norm
of the new spectrum.
\noindent Both  Eq.~\ref{eq:ks} and  Eq.~\ref{eq:sumrules} can be written  as
\begin{equation}
\label{eq:hinteq}
\bar{h}_i=\sum_{\nu=1}^{N_f}{K_h}_{i \nu} {\cal{A}}_{\nu},~~~
{K_h}_{i \nu}=
\left\{
\begin{array}{ll}
{K}_{i \nu} & ~ i \le L \\
\frac{1}{B} &  ~ i = L+1 \\
\end{array}
\right. ~~.
\end{equation}
\noindent This is the  basic equation which relates  $h$ to  ${\cal{A}}$ and
determines the  likelihood function $P[h|A] \equiv P[h|{\cal{A}}]$.  MEM will
produce the most probable spectrum ${\cal{A}}$ normalized to $\bar{s}$ which
minimizes the $\chi^2$ function in Eq.~(\ref{eq:chisqn}) subject to the entropy
constraint.

\subsection{Discussion}
\label{ssec:diss}

We want to point that for the one-particle case, where $B=1$,
Eq.~(\ref{eq:sumrules}) is equivalent to
\begin{equation}
\label{eq:gs0gsb}
Gs(0)+Gs(\beta)=s~.
\end{equation}
\noindent By using Eq.~(\ref{eq:sumrules}) in the calculation of the likelihood
function we impose
\begin{equation}
\label{eq:g0gb}
G(0)+G(\beta)=\frac{Gs(0)}{s}+\frac{Gs(\beta)}{s}=1,
\end{equation}
\noindent at every measurement. Since Eq.~\ref{eq:g0gb}  results solely from
the anticommutation relation of the one-particle operators it should be
satisfied in every possible configuration and implicitly in every measurement.
Therefore, this way of implementing the normalization sum-rule is more natural
than the usual way based on Lagrange multipliers where the constraint is
globally imposed, i.e. not at every measurement but only for the final Green's
function obtained at the end of the QMC process.

For the two-particle case, where $B=\chi(T)$, the sum-rule
Eq.~(\ref{eq:sumruleg}) is not an independent equation as in the one-particle
case, but merely an integration over $\tau$ of Eq.~\ref{eq:inteq}. Therefore it
is essential to treat $B$ as a constant (equal to the final, averaged over all
QMC configurations, $\bar{\chi}(T)$) and to disregard measurement dependent
fluctuations in $\chi(T)$. This way we relate the norm of  ${\cal{A}}$ only to
the fluctuation of the sign $s$.

\section{Comparison of the spectra obtained with the two methods}
\label{sec:newdata}

\begin{figure}[t]
\centerline{
\includegraphics*[width=3.3in]{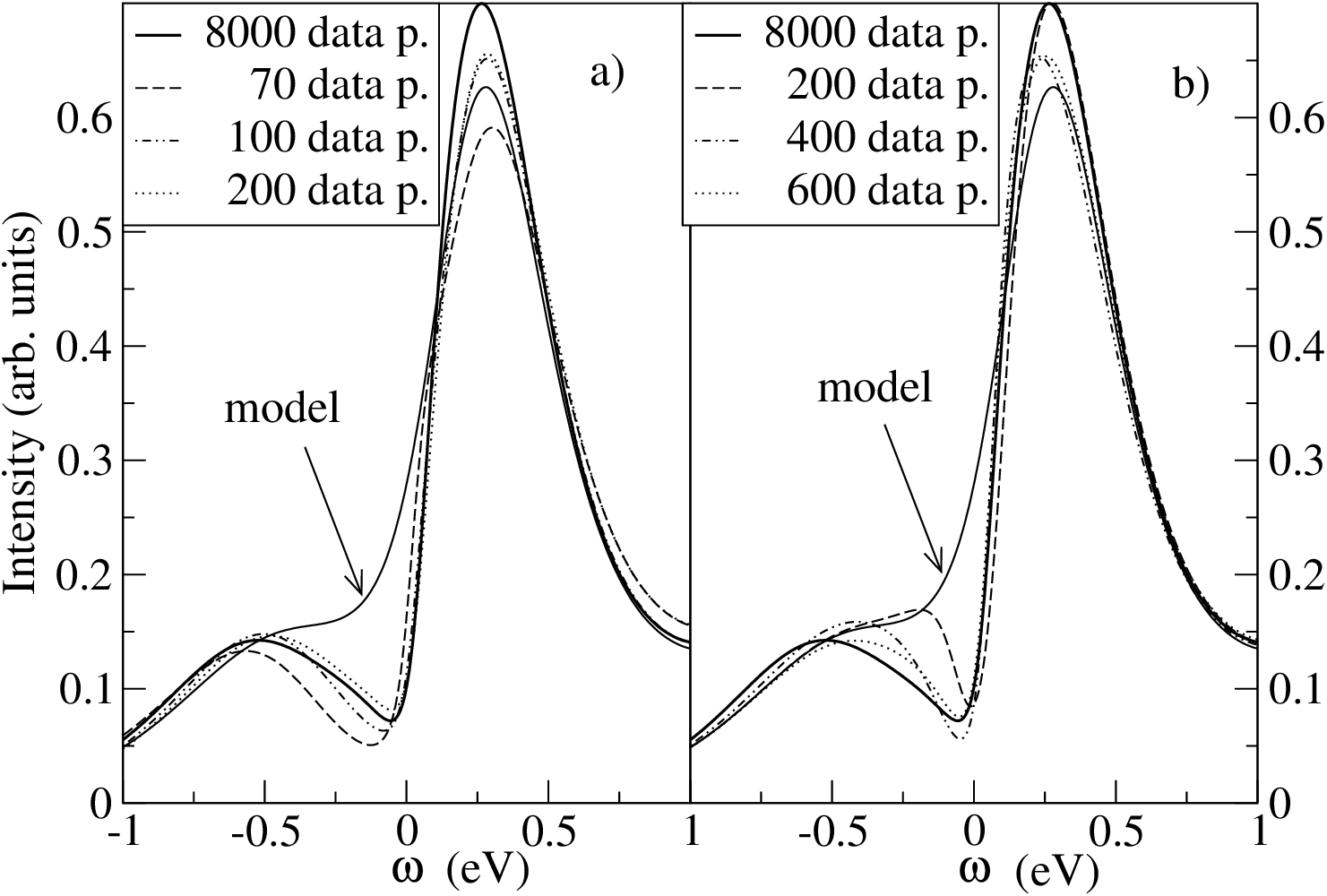}}
\caption{One-particle spectra at $K=(\pi,\pi/2)$ calculated with
different amounts of data using a) the new method and  b) the old
method.}
\label{fig:spectno}
\end{figure}

In this section we present a comparison between the spectra obtained with the
old approach which does not consider the sign covariance, and the new one
described in Sec.~\ref{sec:newmem}.  For calculating the one-particle spectrum
at the highest temperatures, the model $m(\omega)$ used in the entropy
functional Eq.~(\ref{eq:entropy}), is chosen to be a Gaussian function. The
model for lower temperatures is taken to be the spectrum obtained at a slightly
higher temperature, a procedure called annealing.

In  Fig.~\ref{fig:spectno} (a) and (b)  we show the  one-particle
spectra of the Hubbard model at $K=(\pi,\pi/2)$ calculated for different
amounts of data with the new  and respectively with the old method.  In
both cases, when  a large amount of data is used ($8000$ data points)
the spectrum  (thick continuous line) is converged. Moreover the two
methods produce the same spectrum. However, it can be noticed that
with the new method a reasonably good spectrum, i.e. a spectrum close to
the converged one, can be obtained with an amount of data as small as
$100$ data points (see the double-dotted dashed line in
Fig.~\ref{fig:spectno} (a)). On the other hand, the old method requires
at least $600$ data points for a spectrum of comparable quality (see the
dotted  line in Fig.~\ref{fig:spectno} (b)).  Thus in our case we find
that the new method reduces the computational cost of calculating the
one-particle spectra about six times.

In general the calculation of the two-particle spectra turns out to be
more difficult, because the data are more correlated and because a good
default model is missing.  In our case the amount of data needed for
calculating the two-particle spectra  is about one order of magnitude
larger. At high temperature we choose a default model of the form
\begin{equation}
\label{eq:model2}
m(\omega)=\exp [\lambda_0+\lambda_1 \omega \coth(\beta \omega /2)]
\end{equation}
\noindent where $\lambda_0$ and  $\lambda_1$ are
Lagrange multipliers chosen to satisfy certain moment constraints, as
described in ref~\cite{mem}. Again, the annealing technique is used for
lower temperature calculations.

\begin{figure}[t]
\centerline{
\includegraphics*[width=3.3in]{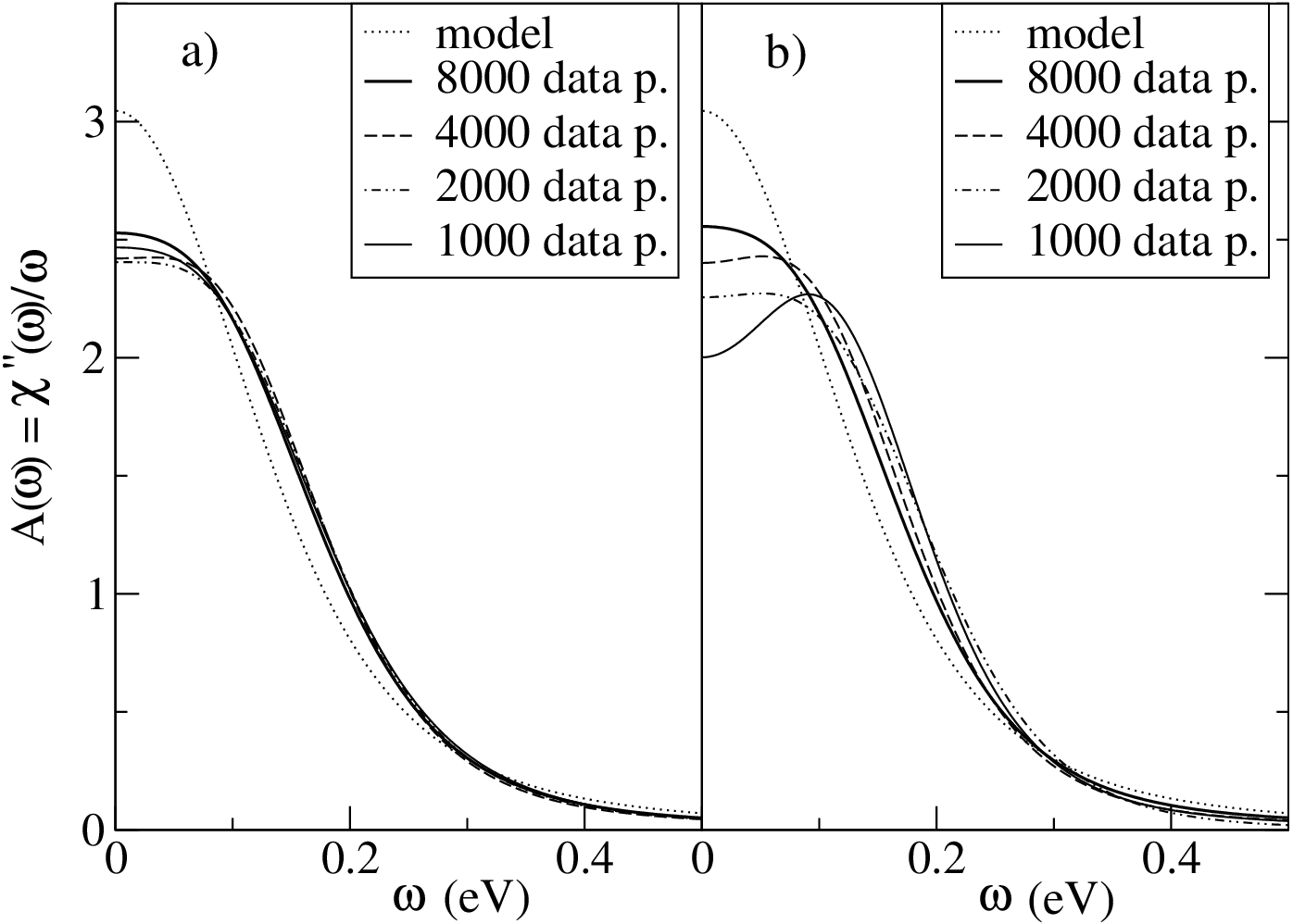}}
\caption{Two-particle spin susceptibility spectra $A(\omega)=\chi''(\omega)/\omega$ at
$K=(0,\pi/2)$ calculated for different amounts of data with a) the new method and
b) the old method.}
\label{fig:chiwcomp}
\end{figure}

\begin{figure}[t]
\centerline{
\includegraphics*[width=3.3in]{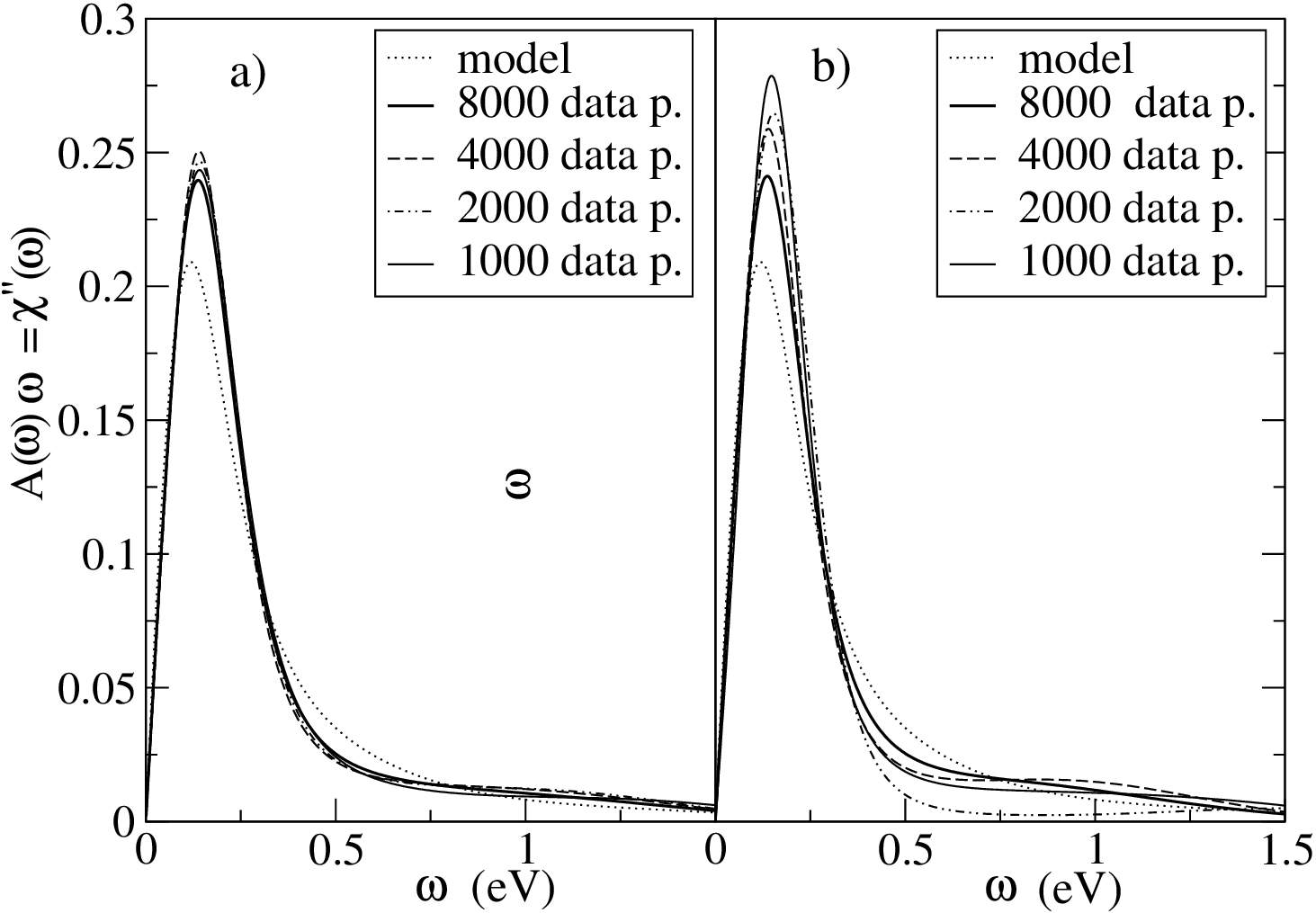}}
\caption{Imaginary part of  the spin susceptibility  $\chi''(\omega)=A(\omega) \omega$ at
$K=(0,\pi/2)$ calculated for different amount of data with a) the new method and
b) the old method.}
\label{fig:chicomp}
\end{figure}

The  spin susceptibility spectra at  $K=(0,\pi/2)$  calculated with the
two methods for different amounts of data is shown in
Fig.~\ref{fig:chiwcomp}.  For the two-particle case the spectra $A(K,
\omega)$ defined in Eq.~\ref{eq:inteq} is in fact $\chi''(K,
\omega)/\omega$ where $\chi''(K, \omega)$ is the imaginary part of the
spin  susceptibility.  When a large amount of data   is used ($8000$
data points) both methods produce the same spectrum (thick continuous
line in both Fig.~\ref{fig:chiwcomp} (a) and (b)). However, for small
amounts of data, the new method produces significantly superior results
to the old method. For example the spectrum obtained with the new method
for $1000$ data points (thin continuous line in Fig.~\ref{fig:chiwcomp}
(a)) is closer to the converged spectrum than the one obtained  with the
old method for  $4000$ data points (dashed line in
Fig.~\ref{fig:chiwcomp} (b)).
The same conclusion can be drawn by comparing the high energy features
($\approx 0.5 -1.5 ~ eV$) visible in the plot of the imaginary part of
the spin susceptibility $\chi''(K, \omega)$ in Fig.~\ref{fig:chicomp}
(a) and (b).

\section{Conclusions}
\label{sec:conc}

We showed that for QMC simulations with a severe sign problem, achieving
a normal distribution of $G$ is extremely difficult. The problem results
from the nonlinear operation which relates $G$ to the measured
quantities $Gs$ and $s$, and from the correlation between the $Gs$
points and $s$.

By absorbing the sign into the definition of the spectrum, the sign
fluctuations will determine the norm of the spectrum. A connection is
thus established between the measured quantities $(Gs,s)$ and the
renormalized spectrum ${\cal{A}}=As$.  The likelihood function is
calculated with regard to the directly measured $(Gs,s)$ data, thus no
nonlinear manipulation of the data is being applied.  The correlations
between  $Gs$ and $s$ can be removed by a rotation in the space
determined by these vectors.

We illustrated the power of this approach by a comparison of the spectra
obtained with the old and the new method. When the sign is small and the
correlation between the sign $s$ and the measured $Gs$ points is
significant, the old method requires a very large amount of measurements
per bin in order to produce the normally distributed and uncorrelated
data points necessary for obtaining good spectra.  In contrast, the new
method provides good spectra for a much smaller amount of data. In our
case  the old method needs about six times more data than the new one,
but for other problems characterized by stronger correlations this
amount can be much larger.

\section*{Acknowledgments} This research was supported by the NSF grants
DMR-0312680 and DMR-0113574. This research used resources of the Center for
Computational Sciences and was sponsored in part by the offices of Advanced
Scientific Computing Research and Basic Energy Sciences, U.S. Department of
Energy. Oak Ridge National Laboratory, where TM is a Eugene P. Wigner Fellow,
is managed by UT-Battelle, LLC under Contract No.\ DE-AC0500OR22725.

\end{document}